\documentclass[final,english]{bullsrsl}[2022/06/15]

\usepackage[latin1]{inputenc}
\usepackage[T1]{fontenc}

\usepackage{natbib} 
\usepackage{graphicx}

\begin{document}
\title{An automated photometric pipeline for the ILMT data}


\author[affil={1,2}, corresponding]{Bhavya}{Ailawadhi}
\author[affil={3,4}]{Talat}{Akhunov}
\author[affil={5}]{Ermanno}{Borra}
\author[affil={1,6}]{Monalisa}{Dubey}
\author[affil={1,6}]{Naveen}{Dukiya}
\author[affil={7}]{Jiuyang}{Fu}
\author[affil={7}]{Baldeep}{Grewal}
\author[affil={7}]{Paul}{Hickson}
\author[affil={1}]{Brajesh}{Kumar}
\author[affil={1}]{Kuntal}{Misra}
\author[affil={1,2}]{Vibhore}{Negi}
\author[affil={1,8}]{Kumar}{Pranshu}
\author[affil={7}]{Ethen}{Sun}
\author[affil={9}]{Jean}{Surdej}
\affiliation[1]{Aryabhatta Research Institute of Observational sciencES (ARIES), Manora Peak, Nainital, 263001, India}
\affiliation[2]{Department of Physics, Deen Dayal Upadhyaya Gorakhpur University, Gorakhpur, 273009, India}
\affiliation[3]{National University of Uzbekistan, Department of Astronomy and Astrophysics, 100174 Tashkent, Uzbekistan}
\affiliation[4]{ Ulugh Beg Astronomical Institute of the Uzbek Academy of Sciences, Astronomicheskaya 33, 100052 Tashkent, Uzbekistan}
\affiliation[5]{Department of Physics, Universit\'{e} Laval, 2325, rue de l'Universit\'{e}, Qu\'{e}bec, G1V 0A6, Canada}
\affiliation[6]{Department of Applied Physics, Mahatma Jyotiba Phule Rohilkhand University, Bareilly, 243006, India}
\affiliation[7]{Department of Physics and Astronomy, University of British Columbia, 6224 Agricultural Road, Vancouver, BC V6T 1Z1, Canada}
\affiliation[8]{Department of Applied Optics and Photonics, University of Calcutta, Kolkata, 700106, India}
\affiliation[9]{Institute of Astrophysics and Geophysics, University of Li\`{e}ge, All\'{e}e du 6 Ao$\hat{\rm u}$t 19c, 4000 Li\`{e}ge, Belgium}

\correspondance{bhavya@aries.res.in}
\date{13th October 2020}
\maketitle


%

\begin{abstract}
The International Liquid Mirror Telescope (ILMT) is a 4-meter survey telescope continuously observing towards the zenith in the SDSS g', r', and i' bands. This survey telescope is designed to detect various astrophysical transients (for example, supernovae) and very faint objects like multiply-imaged quasars and low surface brightness galaxies. A single scan of a 22$'$ strip of sky contains a large amount of photometric information. To process this type of data, it becomes critical to have tools or pipelines that can handle it efficiently and accurately with minimal human biases. We offer a fully automated pipeline generated in Python to perform aperture photometry over the ILMT data acquired with the CCD in Time Delayed Integration (TDI) mode. The instrumental magnitudes are calibrated with respect to the Pan-STARRS-1 catalogue. The light curves generated from the calibrated magnitudes will allows us to characterize the objects as variable stars or rapidly decaying transients.
\end{abstract}


\section{Introduction}
The International Liquid Mirror Telescope (ILMT) \citep{Surdej2018} is the first optical survey telescope in South Asia which carries out multi-band optical imaging of celestial sources passing near the zenith in SDSS g', r', and i' bands. The telescope's primary mirror has a diameter of 4m and is made of a thin film of liquid mercury. \cite{Gibson1991} describes the working principle behind the Liquid Mirror Telescope (LMT), where liquid mercury set in rotation around the vertical axis takes the shape of a paraboloid and forms a mirror. A charge-coupled device (CCD) working in Time Delay Integration (TDI) mode is used to compensate for the motion of the celestial objects due to the Earth's rotation. The effect of star-trail curvature and differential drifts encountered in zenithal observation mode is  mitigated through an optical 5-lens corrector system \citep{Hickson1998}.

LMTs offer a continuous observation along the zenith, enhancing the image quality with lower atmospheric extinction and light pollution. These telescopes are budget friendly and their simple design surpasses that of conventional glass telescopes. They observe a specific region of the sky every night, with a cadence of 24 hours, which facilitates the discovery of transient phenomena, examination of stars with variability periods exceeding a few days, and detection of space debris using image subtraction. A full-fledged image subtraction pipeline including the classification of objects is under development (see the paper by Kumar Pranshu et al.).  The ILMT is a multi-national collaboration between Belgium, Canada and India. The telescope was built by a Belgian company named AMOS. Both the 4m class telescopes installed at the Devasthal Observatory are of mutual interest to Belgian and Indian astronomers and they are working toward smooth operations of the telescopes. The astronomers from these countries are working together to scientifically exploit of the ILMT data highlighting its relevance to the BINA consortium.

The ILMT is equipped with a 4K $\times$ 4K CCD and TDI observation is done by synchronising the parallel charge transfer rate of the CCD with the sidereal rate. Therefore, the exposure time for each object coming into the field of view is 102 seconds. Due to the limited RAM of the acquisition computer, every frame is observed for 102$\times$10 seconds $\sim$ 17 minutes to optimise the readout time. Therefore, the region that the ILMT covers in each frame is 22'$ \times $ 9$ \times $ 22' = 1.21 square degrees, after excluding the first CCD frame recorded during the ramping phase of data acquisition. We have developed a customised pipeline in Python to perform aperture photometry on large-format ILMT images. In this article, the methodology and steps adopted in the development of the pipeline are outlined in Section\,\ref{method}. Some initial results of the pipeline are described in Section\,\ref{results}. The work is summarised in Section\,\ref{summary}.

\section{Methodology}
\label{method}

\begin{figure}[t]
\centering
\includegraphics[width=\textwidth]{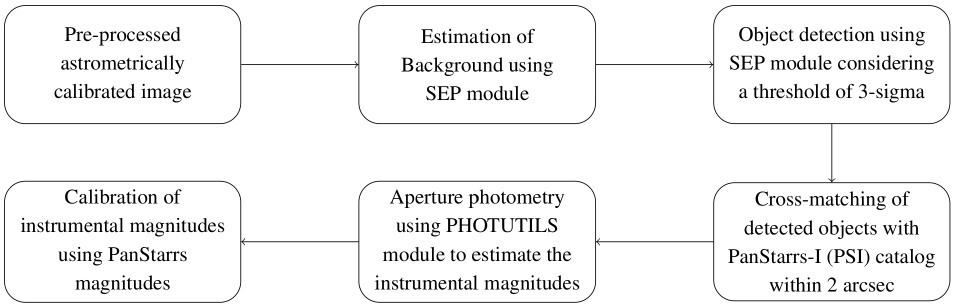}
\smallskip

\begin{minipage}{12cm}
\caption{Schematic diagram of sequence of steps followed for aperture photometry calibration.}
\label{blockdiagram}
\end{minipage}
\end{figure}

This section outlines the steps and techniques used to extract the aperture magnitudes of the point-like sources present on the astrometrically calibrated ILMT images.  The pipeline was tested on data acquired during the first commissioning phase carried out in October--November 2022. The sequence of steps followed is outlined in the flowchart in Fig.\,1 and is described below.

\begin{itemize}
    \item From the raw image, the best dark frame is subtracted to eliminate the effect of dark current. The best dark frame is computed using a median combination of several dark frames, eliminating those with light leaks. Flat fielding is performed to correct for the detector sensitivity variations. This results in a clean pre-processed image.

    \item The Source Extraction and Photometry (\texttt{SEP}) package in Python \citep{Barbary} is used to estimate the background on the image which is then subtracted from the entire image. 
    
    \item A 3-sigma threshold is used in \texttt{SEP} to detect the point sources on the astrometrically calibrated image (Negi et al. 2023). The choice is made to ensure the elimination of a majority of false detections.  
    
    \item A cross-matching algorithm is utilised to match the detected objects with the Pan-STARRS data release 1 (PS1) \citep{Chambers} within a distance of 2 arcsec. This enables us to determine the position and apparent magnitude of the objects, which serve as reference objects for aperture photometry. 
    
    \item Aperture photometry is the measurement of a celestial object's brightness by taking into account the amount of light that enters a circular aperture of 1 Full Width Half Maxima (FWHM) radius centered on the object. Aperture photometry is performed over the matched objects using the  \texttt{PHOTUTILS} \citep{bradley} package in Python. This results in deriving the instrumental magnitudes of the matched objects.
    
    \item Finally, the zero point is evaluated using the instrumental and standard PS1 magnitudes of the matched sources. The instrumental magnitudes of all the stellar sources identified on the ILMT images are calibrated by applying the derived zeropoint.   
\end{itemize}

\section{Initial results from the pipeline}
\label{results}

The steps described in Section\,\ref{method} are integrated into the form of an automated pipeline in Python. The pipeline was tested on three images taken at 04:32 LST on 31/10/2022, 01/11/2022, and 30/10/2022 in the g', r', and i' spectral bands, respectively. These observations were performed during the grey photometric nights and were not affected by the light leak issue. For calibration, we chose stars from PS1 because of the same filters used as those available with the ILMT. The zero point is determined using the instrumental magnitude and PS1 magnitude of the matched sources. After applying the zero point correction, as expected, the calibrated ILMT magnitudes depict a very good agreement with the PS1 magnitudes. 
These results indicate that the dispersion is relatively small in the 13--19 magnitude range whereas it becomes significant beyond 19 magnitude. The number of objects with a deviation less than and greater than 0.5 magnitudes in g', r', and i' bands are mentioned in Table\,\ref{Table1}. The overall good agreement indicates that our photometric pipeline is accurate and reliable. 

\begin{table}
\centering
\begin{minipage}{146mm}
\caption{The number of cross matched sources in the ILMT image with respect to the Panstarrs catalogue and the limiting magnitude are tabulated.}
\end{minipage}
\bigskip

\begin{tabular}{ccccc}
\hline
\textbf{Date} & \textbf{Filter} & \textbf{Matched objects} & \textbf{Matched objects} & \textbf{Limiting magnitude} \\
 & & \textbf{< 0.5 mag} & \textbf{> 0.5 mag} & \textbf{mag}\\
\hline
31/10/2022 & g' & 9141 & 420 & 21.9\\
01/11/2022 & r' & 16359 & 878 & 21.7\\
30/10/2022 & i' & 22758 & 943 & 21.5\\
\hline
\end{tabular}
\label{Table1}
\end{table}

Using the calibrated magnitudes, we estimate the limiting magnitude on the ILMT images which is a measure of the faintest objects detected in these observations. The limiting magnitudes on the CCD frames recorded on  31/10/2022, 01/11/2022, and 30/10/2022 in the g', r', and i' bands are 21.9, 21.7 and 21.5 (Table\,\ref{Table1}) respectively for a 3-sigma detection with an exposure time of 102 s. This information aids in decision-making regarding the filter selection as well as observing conditions so that the signal-to-noise ratio of the images can be enhanced. 

\begin{figure}[t]
\centering
\includegraphics[width=\linewidth]{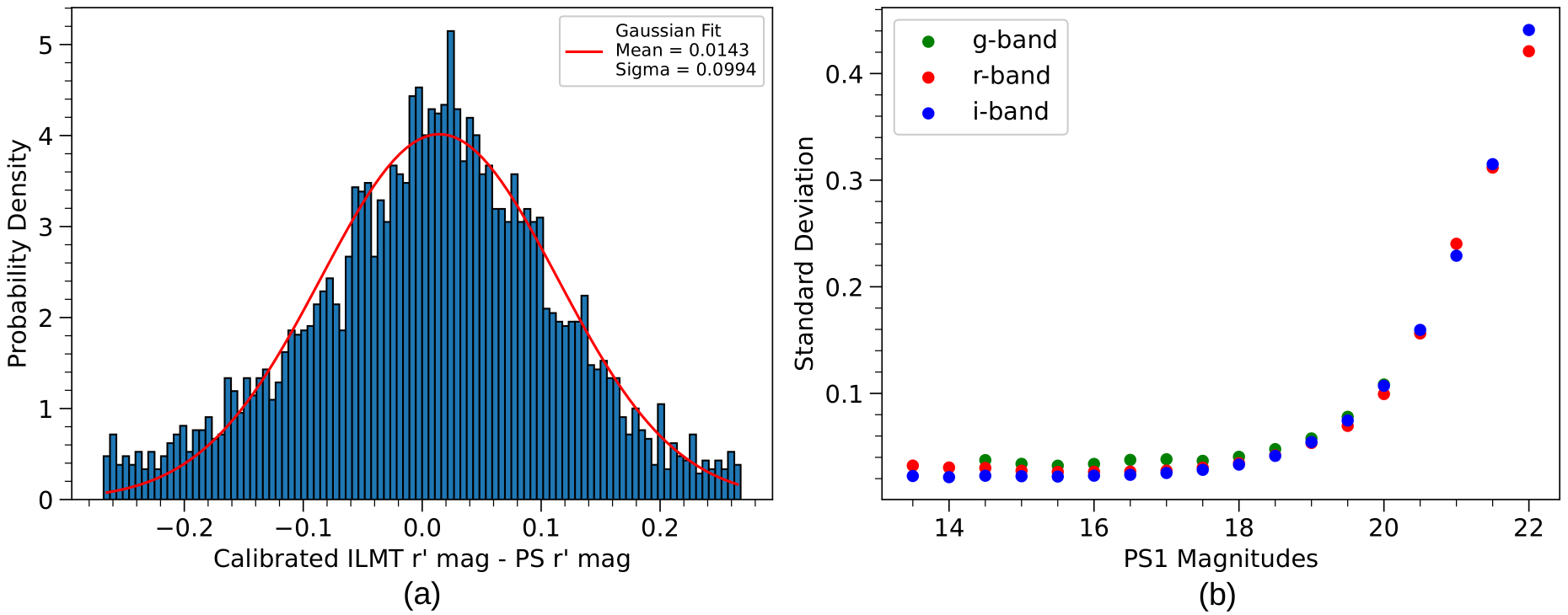}

\begin{minipage}{12cm}
\caption{The left panel (a) shows as an example for the magnitude bin 19.5 < r' < 20.5 the distribution of the differences between the calibrated ILMT r' magnitudes and PS1 r' magnitudes, fitted with a gaussian curve (red curve). The right panel (b) shows the estimated standard deviation in all spectral bands as a function of the PS1 magniutdes.}
\end{minipage}  
\label{deviation}
\end{figure}

The standard deviation as a function of magnitude is evaluated by fitting the Guassian distribution of the differences between calibrated ILMT and PS1 magnitudes lying within a given magnitude bin, i.e. the selected magnitude $\pm$0.5 magnitudes. For consideration, Fig.\,\ref{deviation}(a) depicts the distribution for r' magnitude = 20 lying in the 19.5--20.5 magnitude bin. This is the sigma of the gaussian distribution quoted as the standard deviation. In Fig.\,\ref{deviation}(b), the standard deviation for each magnitude in the three available spectral bands (g', r', and i' bands) is plotted against the PS1 magnitudes. The standard deviation is less than 0.04 for sources with magnitudes in the $[13.5-18]$ range for all the bands. At the fainter end, the deviation increases and reaches up to $\sim$0.4 at 22 magnitude. Since there are fewer fainter objects detected in the g' spectral band, it is hard to evaluate the standard deviation for g'=22 magnitude.

\section{Summary and Conclusions}
\label{summary}

In this work, we present initial results of the automated photometric pipeline, developed in Python, to be used for the ILMT data. The pipeline provides the instrumental magnitudes of all the detected point-like sources and computes the zero-point corrections with respect to the PS1 catalogue and determines the calibrated magnitudes.
The performance of the pipeline was tested by comparing the calibrated ILMT magnitudes with the PS1 magnitudes. These were found to be in good agreement. For the single frames observed on 31/10/2022, 01/11/2022 and 30/10/2022, the limiting magnitude for a 3-sigma detection with an exposure of 102 s in g', r', and i' bands were found to be 21.9, 21.7 and 21.5, respectively. This limit will be further improved by stacking multiple images acquired over a long period of time.

Once routine science observations begin, the pipeline will be used to generate long-term light curves of variable sources with the typical cadence of one day. At present, the pipeline performs aperture photometry but in the near future, the scope of the pipeline will include the task of performing PSF photometry as well.

\begin{acknowledgments}
We thank the referee for reviewing the manuscript and providing constructive comments which has allowed for a better presentation of the results. The 4m International Liquid Mirror Telescope (ILMT) project results from a collaboration between the Institute of Astrophysics and Geophysics (University of Li\`{e}ge, Belgium), the Universities of British Columbia, Laval, Montreal, Toronto, Victoria and York University, and Aryabhatta Research Institute of observational sciencES (ARIES, India). The authors thank Hitesh Kumar, Himanshu Rawat, Khushal Singh and other observing staff for their assistance at the 4m ILMT.  The team acknowledges the contributions of ARIES's past and present scientific, engineering and administrative members in the realisation of the ILMT project. JS wishes to thank Service Public Wallonie, F.R.S.-FNRS (Belgium) and the University of Li\`{e}ge, Belgium for funding the construction of the ILMT. PH acknowledges financial support from the Natural Sciences and Engineering Research Council of Canada, RGPIN-2019-04369. PH and JS thank ARIES for hospitality during their visits to Devasthal. B.A. acknowledges the Council of Scientific $\&$ Industrial Research (CSIR) fellowship award (09/948(0005)/2020-EMR-I) for this work. M.D. acknowledges Innovation in Science Pursuit for Inspired Research (INSPIRE) fellowship award (DST/INSPIRE Fellowship/2020/IF200251) for this work. T.A. thanks Ministry of Higher Education, Science and Innovations of Uzbekistan (grant FZ-20200929344).    
\end{acknowledgments}

\begin{furtherinformation}

\begin{orcids}
\orcid{0000-0003-1637-267X}{Kuntal}{Misra}
\orcid{0000-0002-7005-1976}{Jean}{Surdej}
\end{orcids}



\begin{authorcontributions}
This work results from a long-term collaboration to which all authors have made significant contributions.
\end{authorcontributions}

\begin{conflictsofinterest}
The authors declare no conflict of interest.
\end{conflictsofinterest}

\end{furtherinformation}

\bibliographystyle{bullsrsl-en}

\bibliography{S11-P02_AilawadhiB}

\end{document}